\documentclass[twocolumn]{aastex63}

\newcommand\alfven{Alfv\'{e}n~}

\usepackage{amsmath}
\allowdisplaybreaks

\submitjournal{ApJ}

\shorttitle{\alfven waves}
\shortauthors{Yuan et al.}

\graphicspath{{./}{figures/}}

\begin{document}

\title{\alfven wave mode conversion in pulsar magnetospheres}

\correspondingauthor{Yajie Yuan}
\email{yyuan@flatironinstitute.org}

\author[0000-0002-0108-4774]{Yajie Yuan}
\affiliation{Center for Computational Astrophysics, Flatiron Institute, 162 Fifth Avenue, New York, NY 10010, USA}

\author{Yuri Levin}
\affiliation{Center for Computational Astrophysics, Flatiron Institute, 162 Fifth Avenue, New York, NY 10010, USA}
\affiliation{Physics Department and Columbia Astrophysics Laboratory, Columbia University, 538 West 120th Street, New York, NY 10027}
\affiliation{Department of Physics and Astronomy, Monash University, Clayton VIC 3800, Australia}

\author[0000-0002-9711-9424]{Ashley Bransgrove}
\affiliation{Physics Department and Columbia Astrophysics Laboratory, Columbia University, 538 West 120th Street, New York, NY 10027}

\author{Alexander Philippov}
\affiliation{Center for Computational Astrophysics, Flatiron Institute, 162 Fifth Avenue, New York, NY 10010, USA}

\begin{abstract}

The radio emission anomaly coincident with the 2016 glitch of the Vela pulsar may be caused by a star quake that launches \alfven waves into the magnetosphere, disturbing the original radio emitting region. To quantify the lifetime of the \alfven waves, we investigate a possible energy loss mechanism, the conversion of \alfven waves into fast magnetosonic waves. Using axisymmetric force-free simulations, we follow the propagation of \alfven waves launched from the stellar surface with small amplitude into the closed zone of a force-free dipolar pulsar magnetosphere. We observe mode conversion happening in the ideal force-free regime. The conversion efficiency during the first passage of the \alfven wave through the equator can be large, for waves that reach large amplitudes as they travel away from the star, or propagate on the field lines passing close to the Y-point.
However, the conversion efficiency is reduced due to dephasing on subsequent passages and considerable \alfven power on the closed field lines remains. Thus while some leakage into the fast mode happens, we need detailed understanding of the original quenching in order to say whether mode conversion alone can lead to reactivation of the pulsar on a short timescale.

\end{abstract}


\section{Introduction} \label{sec:intro}
Some pulsars are known to have glitches---occasional, sudden spin up events that interrupt the normal, steady spin down. The first glitch was observed in Vela by \citet{1969Natur.222..228R}, and since then many glitches have been observed in other young pulsars \citep[e.g.,][]{2011MNRAS.414.1679E,2018arXiv180104332M}. Until recently, the timing data around the glitch epoch has been sparse due to observational constraints on major radio telescopes. In a remarkable campaign, Palfreyman and colleagues have used the Mount Pleasant 26-m radio telescope in Hobart, Tasmania and the 30-m telescope in Ceduna, South Australia to time Vela continuously for several years, with the specific purpose of study of its glitches.
In 2016, a glitch event in the Vela pulsar was caught and observed with high time resolution such that single pulses during the glitch were recorded
for the first time \citep{2018Natur.556..219P}. 
Coincident with the glitch, there was an unusually broad pulse, followed by a null pulse, then two pulses with unexpectedly low linear polarization fraction. Subsequent pulses in a 2.6 s interval arrived later than usual pulses.
Since radio emission is believed to be connected to magnetospheric current and pair production \citep[e.g.,][]{2008ApJ...683L..41B,2020PhRvL.124x5101P}, the observed changes suggest that the overall magnetospheric machinery was affected by the glitch.

Glitches are believed to be caused by a sudden transfer of angular momentum from the neutron superfluid to the rest of the star. 
While the star is spun down continuously due to external torques, the rotation of the neutron superfluid is fixed as long as the quantized vortices are pinned to the crustal ion lattice \citep{1975Natur.256...25A} or to superconducting proton flux tubes \citep{1998ApJ...492..267R}. Star quakes have been proposed as a mechanism to simultaneously unpin a multitude of vortices and trigger a glitch \citep[e.g.,][]{1976ApJ...203..213R,1996ApJ...457..844L,2002MNRAS.333..613L,2010ApJ...715L.142E}.
\cite{2020ApJ...897..173B} suggested that the same starquake that triggered the 2016 glitch in Vela, could also dramatically alter the radio emission for a short amount of time. In this scenario, the quake launches \alfven waves into the magnetosphere, and as the waves propagate along magnetic field lines, they may generate local regions with enhanced current density that ignites additional pair production. This would change the pulse profile, and may even quench the radio emission if pair production further away on open field lines causes a backflow which screens the polar gap.
Pair production on closed field lines might also modify the pulse profile, when these pair producing regions are very close to the separatrix.
If \alfven waves on closed field lines keep bouncing back and forth, they may keep producing pairs and influence the radio emission for a long time. 
This should be constrained by the observed duration of the radio pulse disturbance.
In addition, \cite{2020ApJ...897..173B}  predict a weak X-ray burst to accompany the magnetospheric disturbance associated with the 2016 Vela glitch. The duration of the burst should be comparable to the dissipation timescale of Alfven waves in the closed magnetopshere. In this paper we study one of the mechanisms of this dissipation.

The energy of \alfven waves may be removed through several channels. Firstly, in the closed zone, as waves bounce back and forth, counter-propagating \alfven waves lead to a turbulent cascade, and energy is dissipated on small scales \citep[e.g.,][]{2019ApJ...881...13L}.
Small scale \alfven waves can also be more efficiently dissipated by Landau damping \citep{1986ApJ...302..120A}.
Secondly, some wave energy may be absorbed by the crust \citep{2015ApJ...815...25L}. Thirdly, \alfven wave packets propogating along dipole field lines become increasingly oblique and dephased, leading to enhanced current density carried by the wave packet. If there is not enough $e^{\pm}$ in the magnetosphere to conduct the current, dissipation may happen through pair production or diffusion of the wave front \citep{2020ApJ...897..173B}. The charge starvation may also cause \alfven waves to convert to electromagnetic modes.
Fourthly, \alfven waves could convert to fast magnetosonic waves in a plasma filled magnetosphere; the latter is not confined to the field lines and can escape from the magnetosphere. In this paper, we focus on the fourth channel, and quantify the efficiency of \alfven waves converting to fast waves in a plasma filled, dipolar magnetosphere in the force-free regime.

The paper is organized as follows. In \S\ref{sec:method} we describe our numerical method and setup. We show the results in \S\ref{sec:non-rotating} and \S\ref{sec:rotating}, for a non-rotating dipolar magnetosphere and a rotating force-free magnetosphere, respectively. We apply the results to the Vela pulsar in \S\ref{sec:Vela}, and conclude with more discussion in \S\ref{sec:conclusion}.

\section{Force-free formalism and numerical method}\label{sec:method}
In a plasma filled pulsar magnetosphere, the electromagnetic energy is much larger than particle kinetic energy, so force free is a good approximation (except for current sheets).
In this regime, the force balance equation is simply 
\begin{equation}\label{eq:FF_constraint}
    \rho\mathbf{E}+\mathbf{J}\times\mathbf{B}=0,
\end{equation}
and the evolution of the electromagnetic field is governed by the following equations \citep[e.g.,][]{1999astro.ph..2288G,2002luml.conf..381B}
\begin{align}
  \frac{\partial\mathbf{E}}{\partial t}&= \nabla\times\mathbf{B}-\mathbf{J},\label{eq:FF_dEdt}\\
  \frac{\partial\mathbf{B}}{\partial t}&=- \nabla\times\mathbf{E},\label{eq:FF_dBdt}\\
  \mathbf{J}&=\nabla\cdot\mathbf{E}\frac{\mathbf{E}\times\mathbf{B}}{B^2}+\frac{(\mathbf{B}\cdot\nabla\times\mathbf{B}-\mathbf{E}\cdot\nabla\times\mathbf{E})\mathbf{B}}{B^2},\label{eq:FF_J}
\end{align}
with the constraints $\mathbf{E}\cdot\mathbf{B}=0$ and $E<B$ (we employ
Heaviside-Lorentz units and set $c=1$). For simplicity, we only consider axisymmetric magnetospheres and axisymmetric perturbations in this work. We first numerically obtain the steady state of a force-free magnetosphere, then launch \alfven waves by applying a small toroidal displacement on the neutron star surface over a small angular range $\theta\in(\theta_1,\theta_2)$. More specifically, we assume a disturbance in the angular velocity of the neutron star surface in the following form during a finite time period $T$:
\begin{equation}\label{eq:perturbation}
    \delta\omega=
    \begin{cases}
    \displaystyle
    \delta\omega_0e^{-\frac{1}{2}\left(\frac{\theta-\theta_m}{\sigma}\right)^2}\sin(2\pi n t/T), & 0\le t\le T,\\
    0, & t>T,
    \end{cases}
\end{equation}
where the Gaussian profile with $\theta_m=(\theta_1+\theta_2)/2$ and $\sigma=|\theta_2-\theta_1|/6$ allows the perturbation to go to zero smoothly at the boundaries $\theta_1$ and $\theta_2$; $n$ is an integer representing the number of wave cycles during time $T$. This generates an  electric field perturbation at the stellar surface
\begin{equation}\label{eq:perturbation_deltaE}
    \delta E_{\theta}=-\delta\omega r_*\sin\theta B_{0r},
\end{equation}
where $r_*$ is the stellar radius and $\mathbf{B}_0$ is the background magnetic field. The magnitude of the magnetic perturbation at the center of the wave packet is 
\begin{equation}
    \left(\frac{\delta B}{B_0}\right)_{r_*}=\delta\omega_0 r_* \sin\theta_m.
\end{equation}
We then follow the subsequent propagation and evolution of the wave packet.

We use our code \emph{Coffee} (COmputational Force FreE Electrodynamics)\footnote{\href{https://github.com/fizban007/CoffeeGPU}{https://github.com/fizban007/CoffeeGPU}} to numerically solve the force-free equations \citep{2020ApJ...893L..38C}. To suit our study of axisymmetric cases, we developed a 2D version using spherical coordinates $(r,\theta)$. 
The basic algorithm is similar to \citet{2015PhRvL.115i5002E, 2016ApJ...817...89Z}: we use fourth-order central finite difference stencils on a uniform $(\log r, \theta)$ grid and a five-stage fourth-order low storage Runge-Kutta scheme for time evolution \citep{carpenter_fourth-order_1994}. We use hyperbolic divergence cleaning \citep{2002JCoPh.175..645D} to enforce $\nabla\cdot\mathbf{B}=0$ so that the error is advected away at $c$ and damped at the same time. 
To enforce the force-free condition, we explicitly remove any
$\mathbf{E}_{\parallel}$ by setting
$\mathbf{E}\to\mathbf{E}-(\mathbf{E}\cdot\mathbf{B})\mathbf{B}/B^2$ at
every time step \footnote{The $\mathbf{E}_{\parallel}$ cleaning is done in addition to evaluating parallel force-free current, not to replace it as in \citet{2006ApJ...648L..51S}.}, and when $E>B$ happens, we reset $\mathbf{E}$ to $(B/E)\mathbf{E}$. 
We apply standard sixth order Kreiss-Oliger numerical
dissipation to all hyperbolic variables to suppress high frequency noise
from truncation error \citep{kreiss_methods_1973}. At the outer boundary, we implement an
absorbing layer to damp all outgoing electromagnetic waves
\citep[e.g.,][]{2015MNRAS.448..606C,2019MNRAS.487.4114Y}. The code is
parallelized and optimized to run on GPUs as well as CPUs with excellent scaling.

Our simulation grid for runs in \S\ref{sec:non-rotating} has 3360 cells equally spaced in $\log r$ between $r=e^{-0.2}r_*=0.82r_*$ and $r=e^{4.8}r_*=121.51r_*$ (absorbing layer is not used), and 2048 cells uniformly distributed in $\theta\in(0,\pi)$. For runs in \S\ref{sec:rotating}, the simulation grid has 4096 cells in $\theta$ direction, and 7680 cells in $\log r$ direction between $r=e^{-0.2}r_*=0.82r_*$ and $r=e^{5.7}r_*=298.87r_*$, within which the last 15 cells are absorbing layers.

\section{Alfv\'{e}n waves in a non-rotating dipole field}\label{sec:non-rotating}
Let us first consider a non-rotating dipole. The magnetic field is purely poloidal, and can be written as 
\begin{equation}
    \mathbf{B}_0=\frac{\nabla\psi\times\hat{\phi}}{r\sin\theta},
\end{equation}
where $\psi=\mu\sin^2\theta/r$ is the flux function, $\mu$ is the magnetic dipole moment, and $\hat{\phi}$ is the unit vector along azimuthal direction. Magnetic field lines lie on constant $\psi$ surfaces; they are described by the equation
\begin{equation}
    r=r_{\rm eq}\sin^2\theta,
\end{equation}
where $r_{\rm eq}$ is the radius where the field line intersects the equatorial plane.

Under axisymmetry constraint, the wave vectors need to be purely poloidal. As a result, \alfven waves have toroidal $\delta\mathbf{B}$ and poloidal $\delta\mathbf{E}$, while fast modes have toroidal $\delta\mathbf{E}$ and poloidal $\delta\mathbf{B}$. Therefore, the two modes are easily distinguished by their polarizations.

\begin{figure}
    \centering
    \includegraphics[width=0.95\columnwidth]{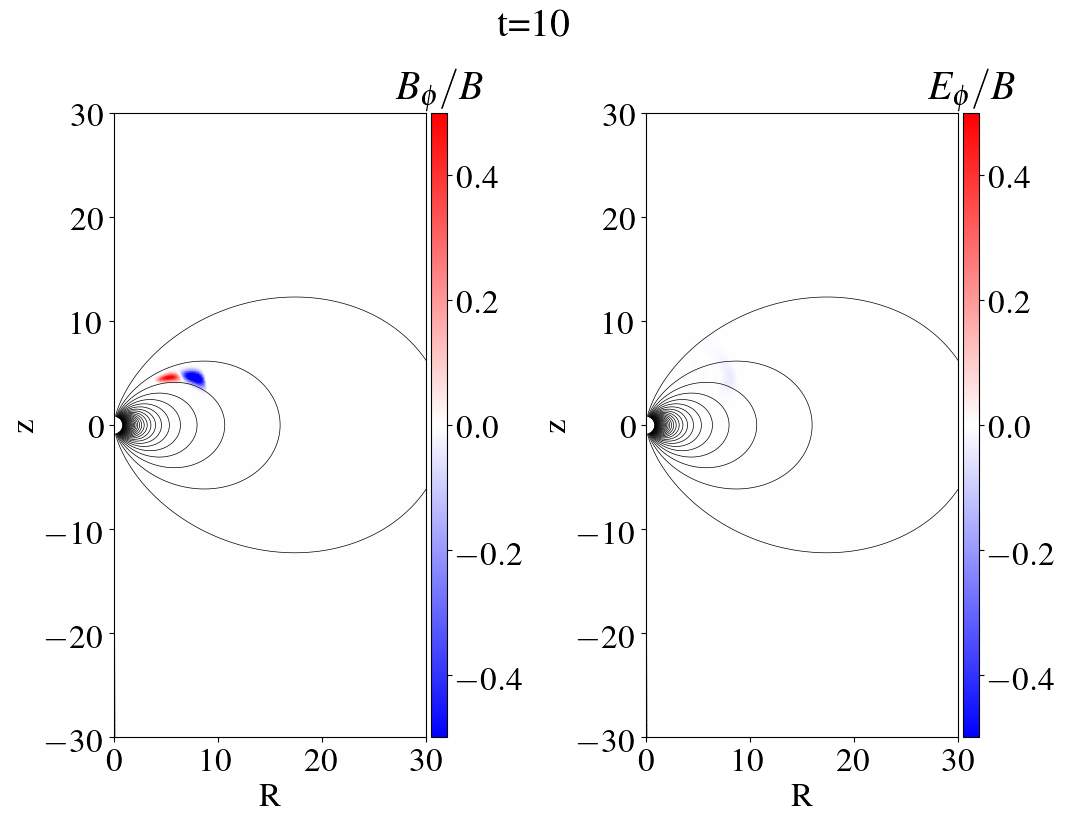}\\
    \includegraphics[width=0.95\columnwidth]{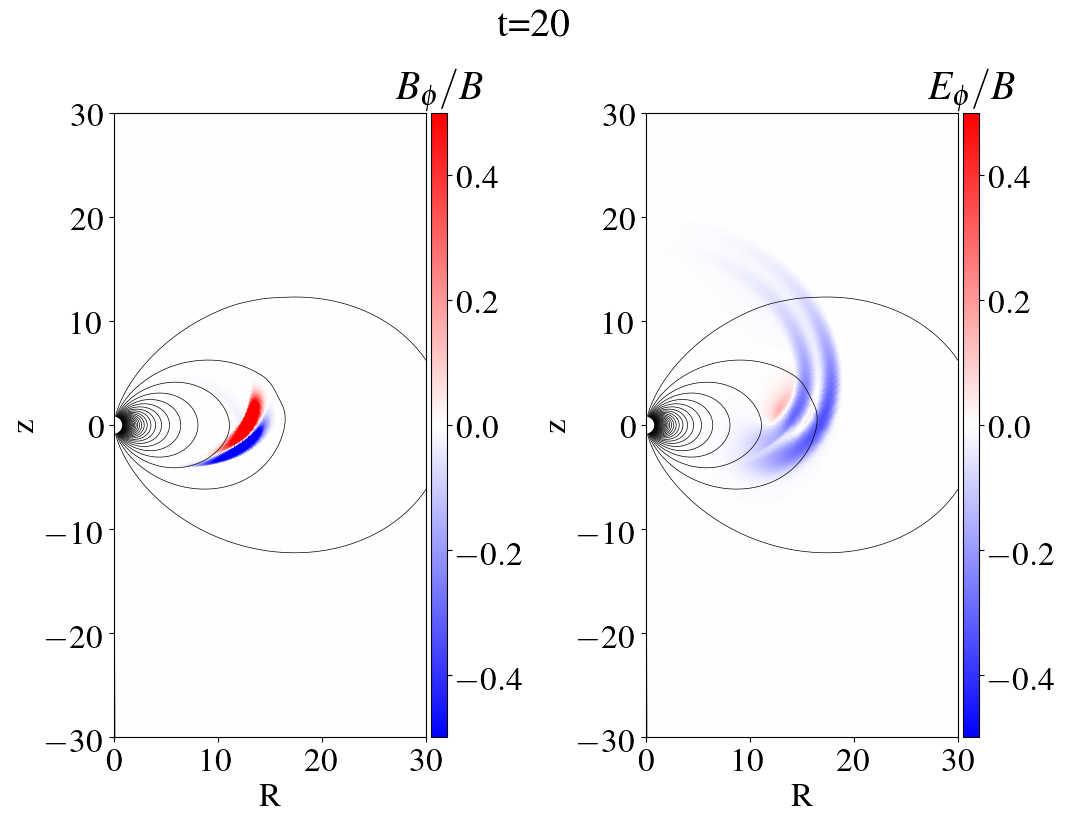}\\
    \includegraphics[width=0.95\columnwidth]{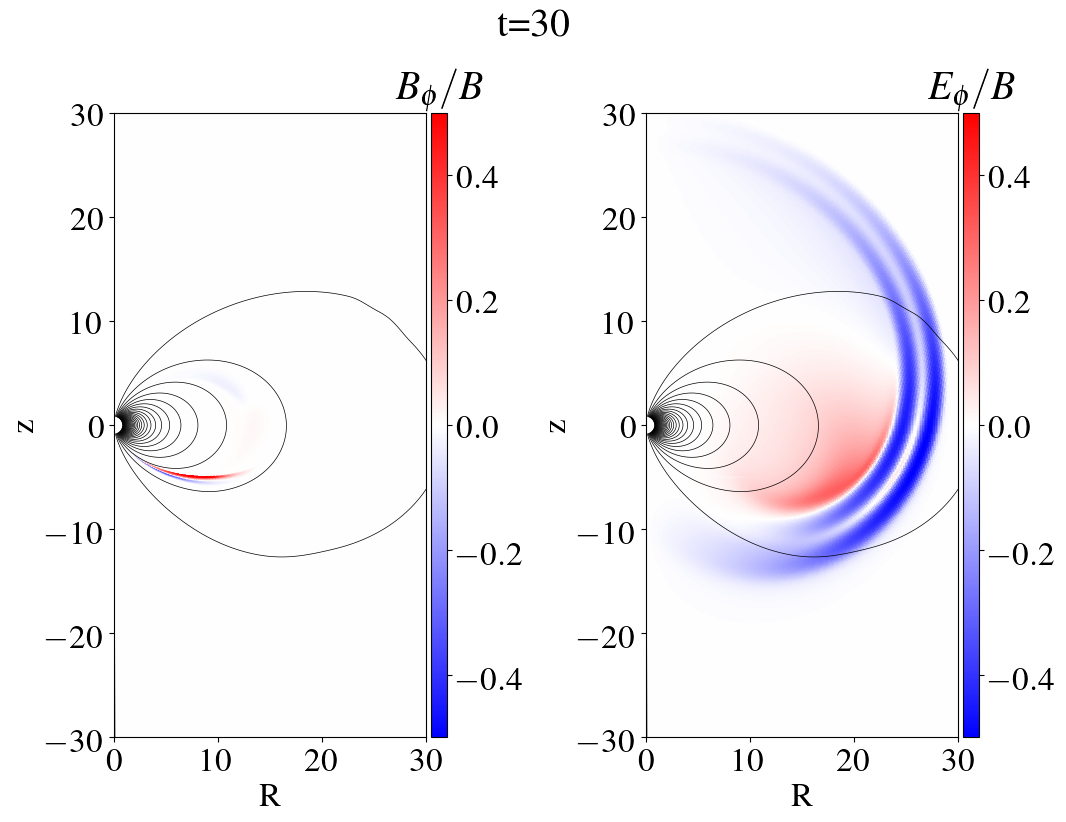}
    \caption{Snapshots of wave field evolution in a non-rotating dipole field. In this example, the initial Alfv\'{e}nic perturbation has a duration of $T=5r_*/c$ with only one full cycle, and is launched inside the flux tube whose equatorial intersection is bounded by $r_{\rm eq}=10r_*$ and $15r_*$; the center of the wave packet passes through $r_m=12.1r_*$. From top to bottom three different time slices are shown. Left panels show $B_{\phi}/B$, manifestation of \alfven modes; right panels show $E_{\phi}/B$, manifestation of fast modes. In the plot, lengths are in units of $r_*$ and time is in units of $r_*/c$ (same below).}
    \label{fig:nonrotating-fields}
\end{figure}

\begin{figure}
    \centering
    \includegraphics[width=\columnwidth]{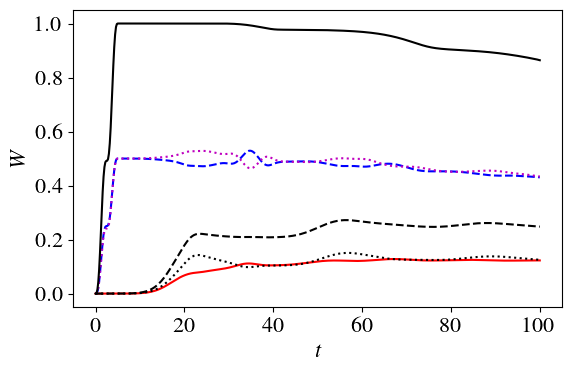}
    \caption{Energy evolution as a function of time for different wave components, corresponding to the example shown in Figure \ref{fig:nonrotating-fields}. Blue dashed line: total magnetic energy in all the wave components; magenta dotted line: total electric energy in all the wave components; black solid line: total electromagnetic energy in all the wave components; red solid line: electric energy of the fast mode; black dotted line: magnetic energy of the fast mode; black dashed line: total electromagnetic energy in the fast mode. All values have been normalized to the initial injected energy $W_0$.}
    \label{fig:nonrotating-energy}
\end{figure}

\begin{figure}
    \centering
    \includegraphics[width=\columnwidth]{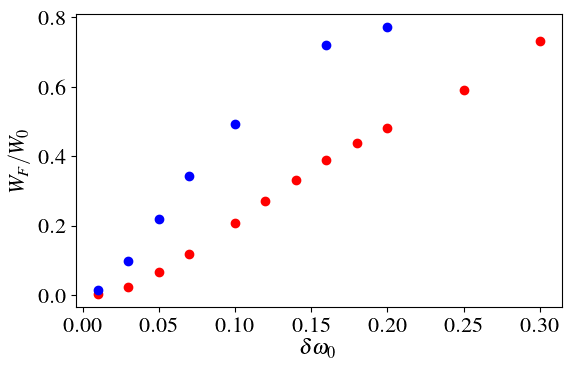}\\
    \includegraphics[width=\columnwidth]{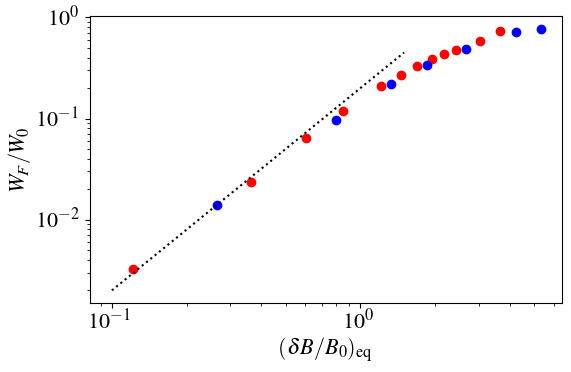}
    \caption{Top: efficiency of converting to fast mode $W_F/W_0$ after the first passage of the \alfven wave through the equator, in a non-rotating dipole background field. Horizontal axis is the initial perturbation magnitude $\delta\omega_0$. Red points correspond to waves launched on a flux tube with $r_m=12.1r_*$; blue points correspond to waves launched on a flux tube with $r_m=26.6r_*$. Bottom: the same conversion efficiency, plotted against $(\delta B/B_0)_{\rm eq}$, the theoretical \alfven wave amplitude at the equator. The dashed line has the expression $W_F/W_0=0.2(\delta B/B)_{\rm eq}^2$.}
    \label{fig:nonrotating-eff_dw_dB}
\end{figure}

Figure \ref{fig:nonrotating-fields} shows one example of an \alfven wave packet propagating out along the dipole field lines. The group velocity of the wave is $c$ and it is directed along the background magnetic field.  For small amplitude \alfven waves, energy conservation implies that $\delta B^2 A=const.$, where $A$ is the cross sectional area of the flux tube in which \alfven wave is launched. Since the poloidal magnetic flux of the background field is conserved, $B_0 A=const.$, we have $\delta B\propto B_0^{1/2}\propto r^{-3/2}$, and $\delta B/B_0\propto r^{3/2}$, namely, the relative amplitude of the wave grows as it propagates to large radius. Conversion to fast mode becomes significant when $\delta B/B_0$ gets large, and peaks near the equator where $\delta B/B_0$ is largest.
This can be understood qualitatively from the following picture: the launched alfven wave is initially guided purely by magnetic tension, but when $\delta B / B$ approaches 1 the pressure of the perturbation $\delta B^2 \sim B^2$ can deform the background poloidal field, launching a wave driven by magnetic pressure and tension (fast mode).

The total wave energy can be calculated from (Appendix \ref{sec:wave_energy})
\begin{equation}\label{eq:wave_energy}
    W=\int\frac{1}{2}(\delta\mathbf{B}^2+\delta \mathbf{E}^2)\,dV,
\end{equation}
and the energy of the fast mode is
\begin{equation}
    W_F=\int\frac{1}{2}(\delta\mathbf{B}_p^2+\delta \mathbf{E}_{\phi}^2)\,dV,
\end{equation}
where $\delta\mathbf{B}_p$ denotes the poloidal components of $\delta\mathbf{B}$, and $\delta \mathbf{E}_{\phi}$ is the toroidal component of $\delta\mathbf{E}$. Figure \ref{fig:nonrotating-energy} shows the time evolution of the wave energies for the example of Figure \ref{fig:nonrotating-fields}. We can see periodic increase in the fast wave energy (black dashed line); this corresponds to each passage of the \alfven wave packet through the equator where most of the fast wave is generated. The total wave energy (black solid line) should in principle be conserved, but we observe stair-like decreases around $t=35r_*/c$ and $t=70r_*/c$.
This is because when the \alfven wave packet propagates back toward the stellar surface, it is strongly dephased \citep{2020ApJ...897..173B}; both the dephasing and the spatial contraction following the dipole field lines lead to wave variation happening on very small scales. Numerical dissipation becomes important when these small scale structures are not well resolved by the grid. We do find the dissipation decrease as we increase the resolution. Conversion to fast mode, on the other hand, does not depend on resolution at all (Appendix \ref{sec:resolution}). Most of the conversion happens on first passage of the \alfven wave through the equator, before the numerical dissipation effect becomes important.

To quantify the efficiency of \alfven waves converting to fast mode, we measure the fast wave energy $W_F$ at the end of the first passage of the \alfven wave through the equator, and compare that with the initially injected \alfven wave energy $W_0$. We carry out a series of experiments by launching \alfven waves with different magnitude and on different flux tubes. The top panel of Figure \ref{fig:nonrotating-eff_dw_dB} shows the measured conversion efficiency, plotted against the initial perturbation magnitude. The two trends correspond to waves on two different flux tubes. When we instead plot the efficiency against the theoretically computed \alfven wave amplitude at the equator, $(\delta B/B_0)_{\rm eq}=(\delta B/B_0)_{r_*}(r_m/r_*)^{3/2}$, where $r_m=r_*/\sin^2\theta_m$ is the radius at which the center of the \alfven wave packet passes through the equator, then all the points lie on one single trend, as shown in the bottom panel of Figure \ref{fig:nonrotating-eff_dw_dB}. 

\begin{figure}
    \centering
    \includegraphics[width=\columnwidth]{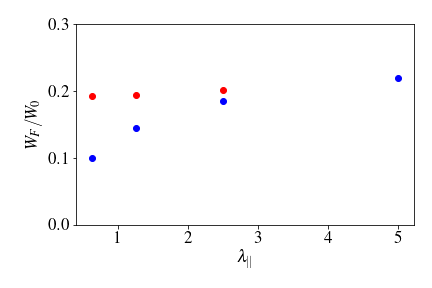}
    \caption{Scaling of the conversion efficiency as a function of the wavelength $\lambda_{\parallel}$, for waves launched on a fixed flux tube with $r_m=12.1r_*$. Blue points correspond to wave packets with a single wavelength, i.e., the perturbation duration $T=\lambda_{\parallel}/c$; the red dots correspond to wave trains with a fixed total duration $T=5r_*/c$.}
    \label{fig:eff_lambda_nonrotating}
\end{figure}

The measured efficiency $W_F/W_0$ has very little dependence on the angular width of the initial \alfven wave perturbation $|\theta_2-\theta_1|$, as long as $|\theta_2-\theta_1|\ll 1$. The conversion efficiency does depend on the total duration $T$ of the \alfven wave perturbation: when $T$ becomes very short, the conversion efficiency drops. In the regime $T\ll r_m/c$, the wave packet has a small length compared to the radius of curvature of the field line, so WKB approximation is applicable. In WKB limit, the wave evolves adiabatically on the \alfven eigenstate; the expected conversion efficiency should go to zero. But for $T\gtrsim0.2r_m$, $W_F/W_0$ only varies slowly with $T$. We also find that the conversion efficiency does not depend on the wavelength $\lambda_{\parallel}=cT/n$ in this case. The scaling of conversion efficiency with $T$ and $\lambda_{\parallel}$ is shown in Figure \ref{fig:eff_lambda_nonrotating}.

The above results suggest that the conversion efficiency only depends on $(\delta B/B_0)_{\rm eq}$ for sufficiently long wave trains. This is essentially a consequence of the self-similarity of the dipole field. Since most of the conversion happens at large radii, especially when the \alfven wave packet passes through the equator, the initial location of wave launch becomes unimportant.

At small \alfven wave amplitude, we find that $W_F/W_0\propto(\delta B/B_0)_{\rm eq}^2$. This is consistent with the three wave interaction theory \citep[e.g.,][]{1998PhRvD..57.3219T,2019MNRAS.483.1731L}. An \alfven wave $A$ can convert to a forward propagating fast mode $F$ and another backward propagating \alfven mode $A_1$ (we can see a small amplitude backward propagating \alfven mode in the bottom row of Figure \ref{fig:nonrotating-fields}).
The amplitude of the fast mode satisfies 
\begin{equation}
    \delta E_{F}\propto\delta E_{A}\delta E_{A_1}.
\end{equation}
Since $\delta E_{A_1}$ is generated by $\delta E_A$ due to propagation along curved field lines, we see that $\delta E_{F}\propto\delta E_{A}^2$. This leads to $W_F/W_0\propto\delta B_A^2$, consistent with the quadratic relation we see in the bottom row of Figure \ref{fig:nonrotating-eff_dw_dB}. A caveat is that theoretical analysis of three-wave interactions is usually carried out in a uniform background magnetic field, thus strictly speaking only applicable when the wavelengths are much smaller than the length scales of field variation. To study relatively large wavelength waves in a dipole field as we do here, numerical simulation is necessary.

At large \alfven wave amplitude $(\delta B/B)_{\rm eq}>1$ the wave interaction becomes highly dynamic, and the result deviates from the above perturbation theory. Some field lines may be opened up, creating a current sheet that eventually dissipates through reconnection. 
When $(\delta B/B)_{\rm eq}>1$ but the energy of the \alfven wave packet $\mathcal{E}_A$ is small compared to the magnetospheric energy $\mathcal{E}_B(r_{\rm eq})$ at $r_{\rm eq}$, only a small portion of the field lines open up near the equator, which then quickly reconnect and relax back. However, when $\mathcal{E}_A>\mathcal{E}_B(r_{\rm eq})$, the \alfven wave packet can break out from the magnetosphere and eject a plasmoid into the pulsar wind. This was recently studied by \cite{2020ApJ...900L..21Y} in the context of fast radio bursts produced by the galactic magnetar 1935+2154 \citep{2020Natur.587...54T,2020Natur.587...59B}.

\section{Alfv\'{e}n waves in a rotating dipole field}\label{sec:rotating}

\begin{figure}
    \centering
    \includegraphics[width=0.7\columnwidth]{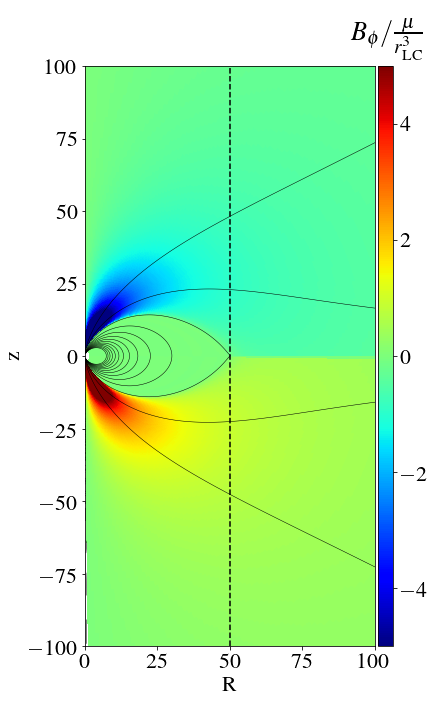}
    \caption{Force-free steady state of a rotating dipolar magnetosphere. Thin solid black lines are poloidal field lines and color represents $B_{\phi}$. The light cylinder is at $50r_*$ (denoted by the vertical dashed line).}
    \label{fig:rotating-steady}
\end{figure}

\begin{figure}
    \centering
    \includegraphics[width=\columnwidth]{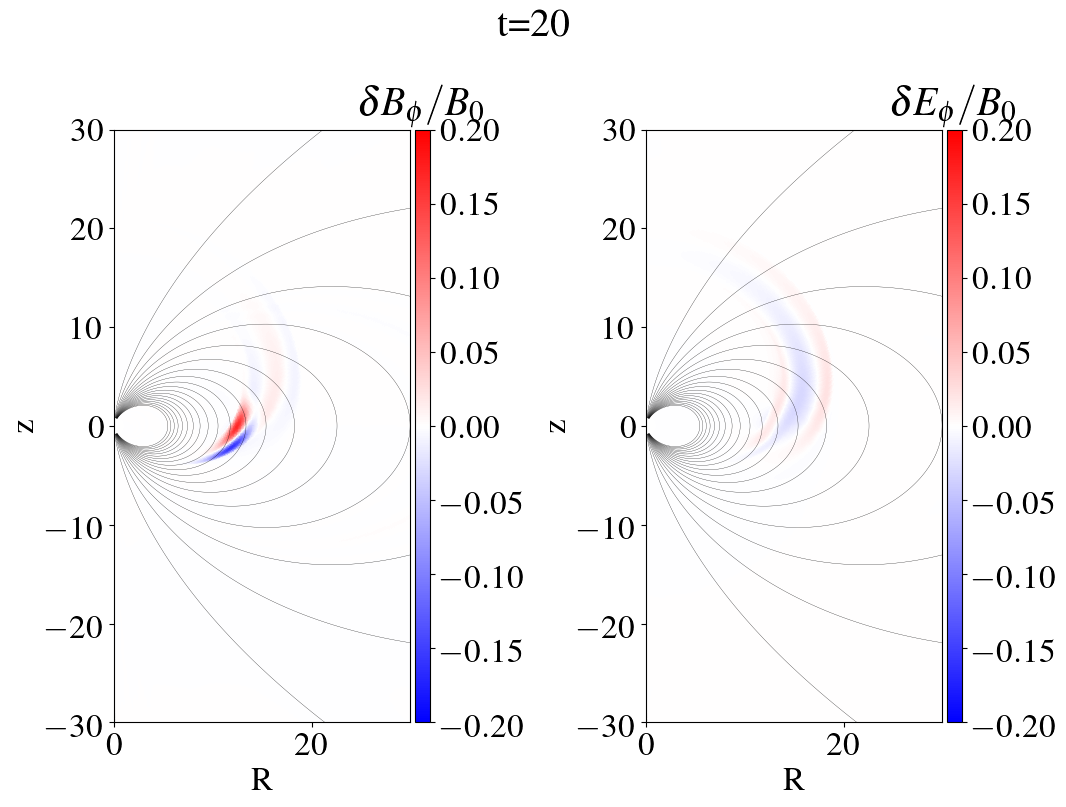}\\
    \includegraphics[width=\columnwidth]{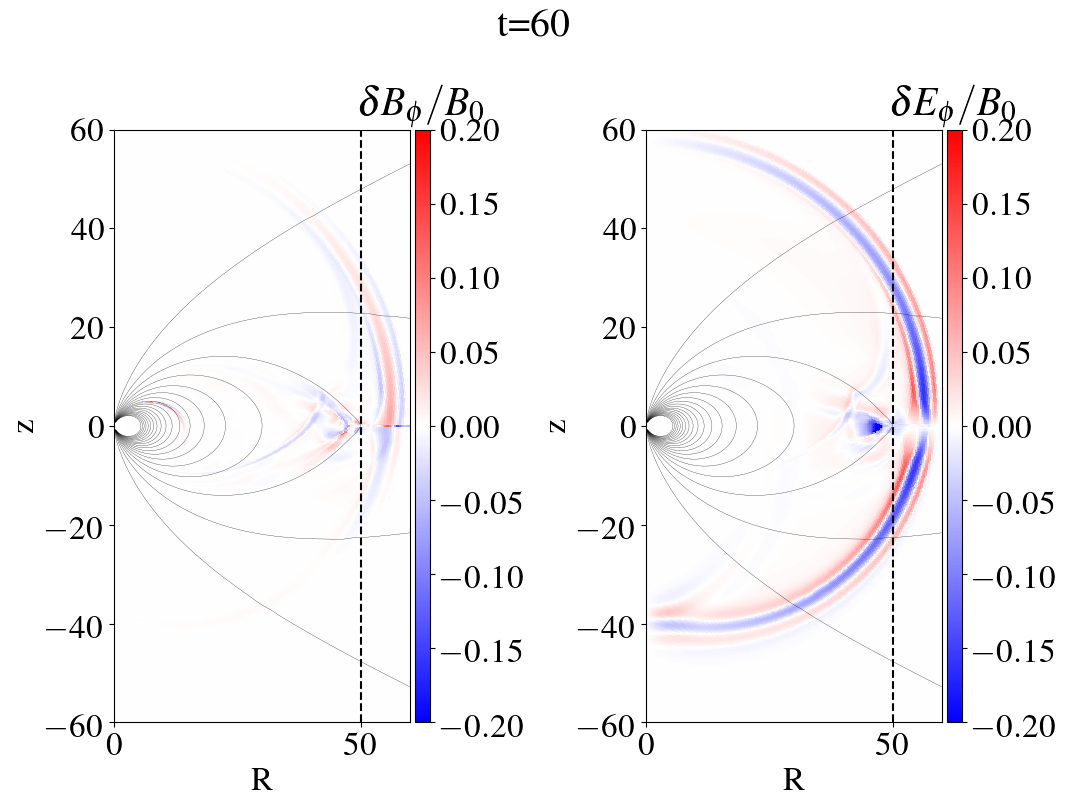}
    \caption{Snapshots of wave field evolution in a rotating dipole field of Figure \ref{fig:rotating-steady}. In this example, the initial Alfv\'{e}nic perturbation has a duration of $T=5r_*/c$ with only one full cycle, and is launched inside the flux tube whose equatorial intersection is bounded by $r_{\rm eq}=10r_*$ and $15r_*$; the center of the wave packet passes through $r_m=12.1r_*$. This is the same flux tube as Figure \ref{fig:nonrotating-fields}. From top to bottom two different time slices are shown. Left panels show $\delta B_{\phi}/B_0$, and right panels show $\delta E_{\phi}/B_0$. Note that the spatial scales are different for the top and bottom panels.}
    \label{fig:rotating-dw0.01r1_10-fields}
\end{figure}

\begin{figure}
    \centering
    \includegraphics[width=\columnwidth]{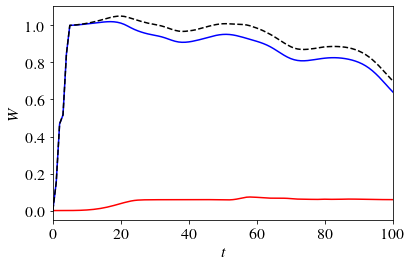}
    \caption{Wave energy evolution for the example shown in Figure \ref{fig:rotating-dw0.01r1_10-fields}. Blue line corresponds to wave energy measured inside the flux surface $\psi=\psi(r_*,\theta_1)$ where the \alfven wave is launched; red line corresponds to wave energy measured outside this flux surface; black dashed line is the sum of the two.}
    \label{fig:rotating-dw0.01r1_10-energy}
\end{figure}

\begin{figure}
    \centering
    \includegraphics[width=\columnwidth]{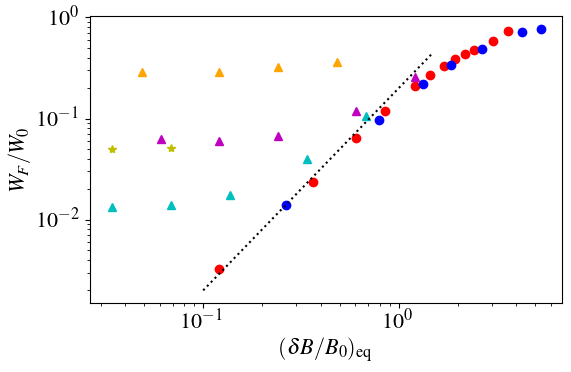}
    \caption{Measured efficiency of \alfven waves converting to fast waves in a rotating dipolar magnetosphere (triangles and stars), plotted together with the non-rotating measurements of Figure \ref{fig:nonrotating-eff_dw_dB} (red and blue dots). The dashed line has the expression $W_F/W_0=0.2(\delta B/B)_{\rm eq}^2$. Cyan triangles are measured for \alfven waves launched on a flux tube with $r_m=6.8r_*$; magenta triangles have $r_m=12.1r_*$; orange triangles have $r_m=24.2r_*$. All three have the same background as shown in Figure \ref{fig:rotating-steady}. Yellow stars are measured for \alfven waves launched on the same flux tube as the cyan triangles, but the background pulsar angular velocity is doubled.}
    \label{fig:eff_dB_log}
\end{figure}

\begin{figure}
    \centering
    \includegraphics[width=\columnwidth]{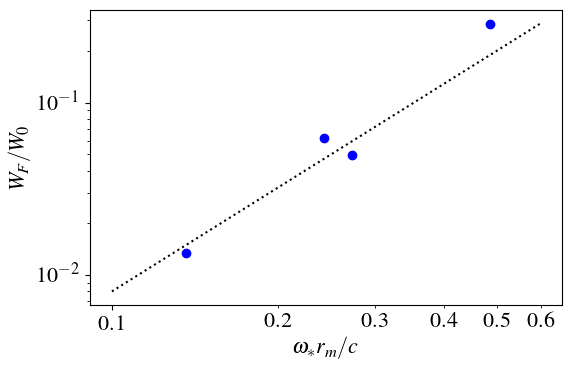}
    \caption{Conversion efficiency of small amplitude \alfven waves, plotted against $\omega_* r_m/c$. The black dashed line has the expression $W_F/W_0=0.8(\omega_*r_m/c)^2$.}
    \label{fig:eff_wrm}
\end{figure}

\begin{figure}
    \centering
    \includegraphics[width=\columnwidth]{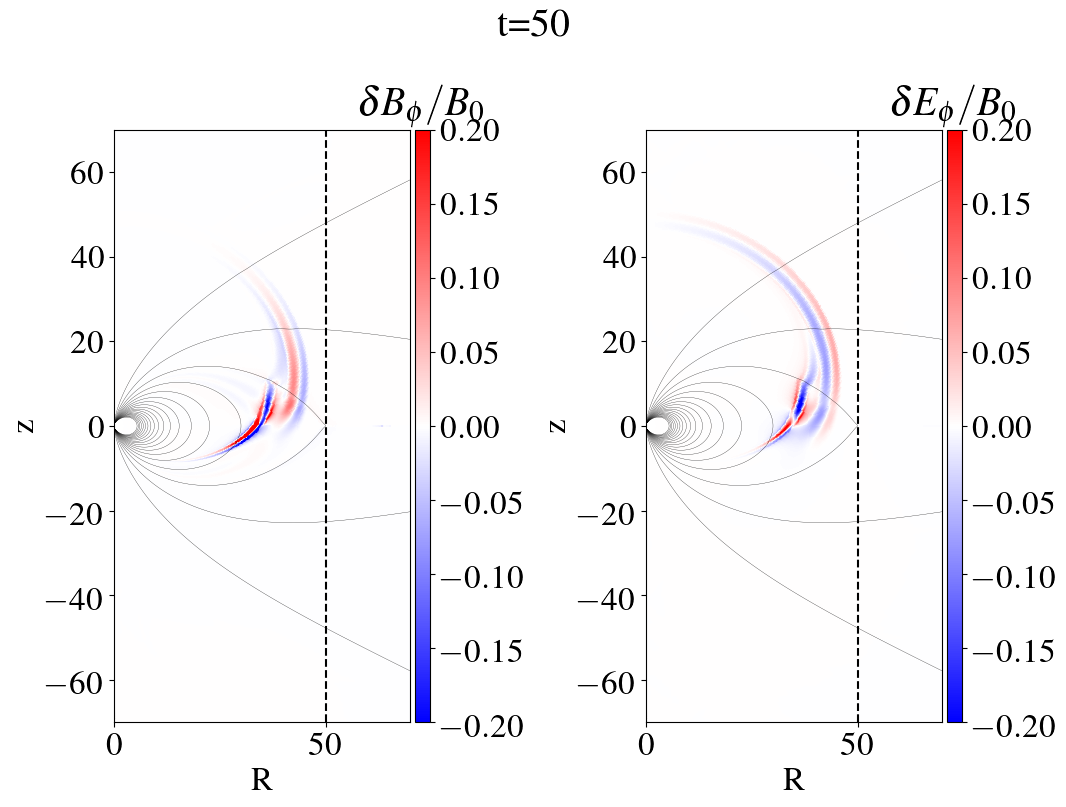}\\
    \includegraphics[width=\columnwidth]{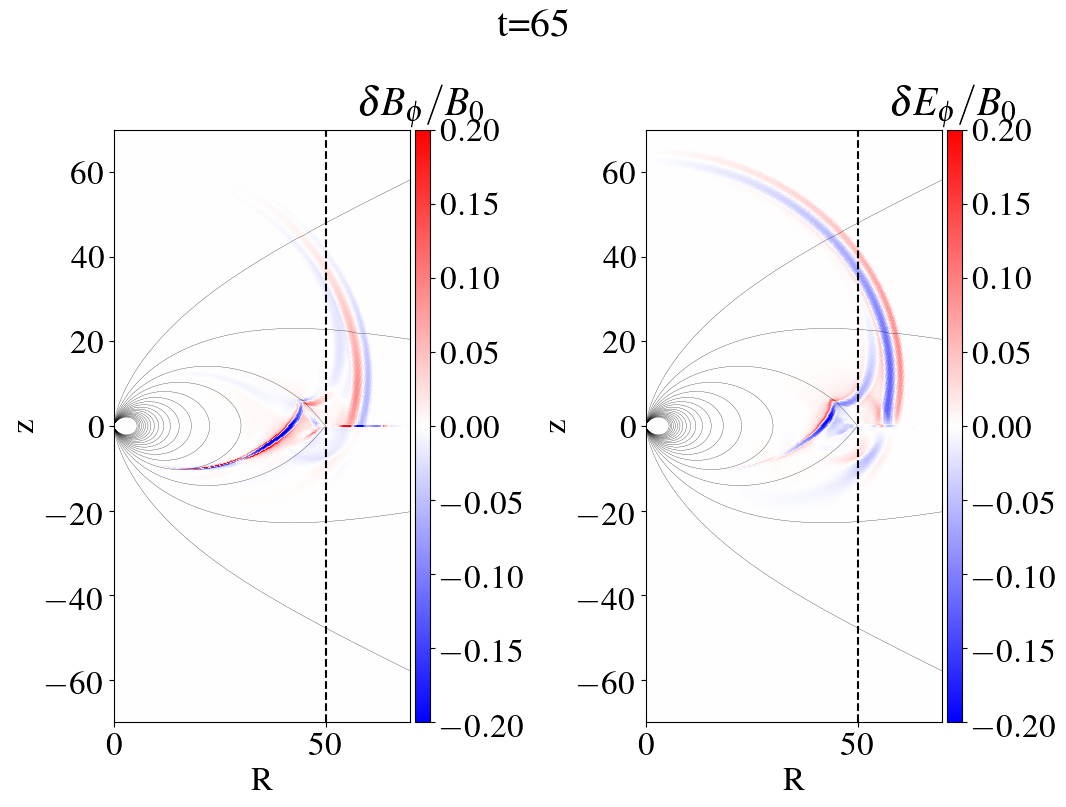}
    \caption{Another example of wave field evolution in the rotating dipole field of Figure \ref{fig:rotating-steady}. In this example, the initial Alfv\'{e}nic perturbation has a duration of $T=5r_*/c$ with only one full cycle, and is launched inside the flux tube whose equatorial intersection is bounded by $r_{\rm eq}=29r_*$ and $43r_*$ in the rotating magnetosphere. From top to bottom two different time slices are shown. Left panels show $\delta B_{\phi}/B_0$, and right panels show $\delta E_{\phi}/B_0$.}
    \label{fig:rotating-dw0.01r1_25-fields}
\end{figure}

\begin{figure}
    \centering
    \includegraphics[width=\columnwidth]{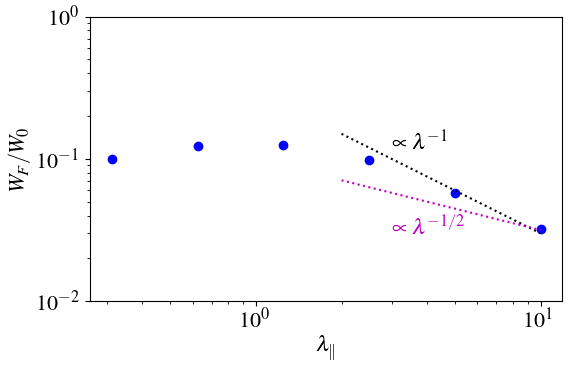}
    \caption{Conversion efficiency as a function of the wavelength $\lambda_{\parallel}$ along the magnetic field line, for small amplitude \alfven waves launched on a flux tube with $r_m=12.1r_*$ in the equilibrium of Figure \ref{fig:rotating-steady}.}
    \label{fig:eff_lambda}
\end{figure}

Now let us consider the case of a rotating magnetosphere. Figure \ref{fig:rotating-steady} shows an example of the steady state field configuration for an aligned dipole rotator. We assume that the light cylinder is located at a radius $r_{\rm LC}=50r_*$. The overall field structure is consistent with e.g., \cite{2006ApJ...648L..51S}. Field lines that go through the light cylinder open up; in this region the magnetic field develops toroidal component and becomes increasingly toroidal at large distances.
A current sheet exists on the equatorial plane outside the light cylinder. Field lines that are closed remain inside the light cylinder; in this region the magnetic field is purely poloidal, but there is an electric field 
\begin{equation}
    \mathbf{E}_0=-(\pmb{\omega}_*\times\mathbf{r})\times\mathbf{B}_0
\end{equation}
that ensures the plasma in the closed zone corotates with the star. The field line separating the closed zone and the open zone is usually called the separatrix; its tip at the light cylinder, where the equatorial current sheet begins, is called the Y point.

Similarly to the non-rotating dipole case, we launch \alfven waves in the closed zone by introducing a small perturbation in the stellar surface angular velocity according to Equation (\ref{eq:perturbation}). Figure \ref{fig:rotating-dw0.01r1_10-fields} shows one example of the wave field evolution. Here the \alfven wave packet is launched on a flux tube that is sufficiently far away from the Y point. 
In the rotating case both the \alfven mode and the fast mode can involve all six $\delta \mathbf{E}$ and $\delta\mathbf{B}$ components. Nevertheless, we find that \alfven mode is dominated by $\delta B_{\phi}$ while fast mode is dominated by $\delta E_{\phi}$, so we plot these field components in Figure \ref{fig:rotating-dw0.01r1_10-fields}. 
The top row of Figure \ref{fig:rotating-dw0.01r1_10-fields} is similar to the middle row of Figure \ref{fig:nonrotating-fields}, except that the wave form of the fast mode is different. As the fast mode propagates toward the Y point, there is increasing interaction with the separatrix and generation of additional \alfven waves near the Y point, as shown in the bottom panels of Figure \ref{fig:rotating-dw0.01r1_10-fields}.


We calculate the wave energy from Equation (\ref{eq:wave_energy}) (see discussion in Appendix \ref{sec:wave_energy}). A complication here is that the \alfven mode and fast mode can no longer be easily separated by their polarizations. However, since the \alfven mode is confined along field lines while the fast mode is not, we can measure the fast mode energy when it has propagated away from the \alfven mode. 
More specifically, the flux surface intersecting the star at $\theta_1$ [namely $\psi=\psi(r_*,\theta_1)$] marks the outer boundary of the flux tube where the \alfven wave is launched.
We measure the wave energy inside and outside this flux surface separately; the energy inside is mostly the \alfven wave energy $W_A$, and the energy outside is mostly the fast wave energy $W_F$.

Figure \ref{fig:rotating-dw0.01r1_10-energy} shows the measured wave energy evolution for the example in Figure \ref{fig:rotating-dw0.01r1_10-fields}. The overall behavior is very similar to the non-rotating case shown in Figure \ref{fig:nonrotating-energy} (the smaller fast wave energy fraction is due to the small perturbation magnitude used in this particular run). 
We can again calculate the efficiency of converting to fast mode from $W_F/W_0$ after the first passage of the \alfven wave packet through the equator, in the same way as before.

Figure \ref{fig:eff_dB_log} shows the measured conversion efficiency for \alfven waves with different magnitude, launched on different flux tubes. When the flux tube is sufficiently far away from the Y point, the deviation of the field from vacuum dipole is small. The center of the flux tube intersects the equator roughly at $r_m=r_*/\sin^2\theta_m$ as before.\footnote{Close to the Y point, the field significantly deviates from vacuum dipole scaling. For example, in the rotating steady state with $r_{\rm LC}=50r_*$, the field lines that originally intersect the equator at $r_m\gtrsim35r_*$ in the non-rotating case all open up.} We 
plot the conversion efficiency $W_F/W_0$ against the theoretical \alfven wave amplitude at the equator $(\delta B/B_0)_{\rm eq}$, similar to Figure \ref{fig:nonrotating-eff_dw_dB}. It is clearly seen that at small \alfven wave amplitude, the scaling deviates from non-rotating dipole cases. The conversion efficiency reaches a constant value, and the value is different for different flux tubes.

We can understand the new scaling by noticing that in the rotating case, the plasma in the closed zone is corotating with the star. There is a spatially varying electric field induced by the rotation, and the \alfven wave can directly interact with this varying background, as shown in Appendix \ref{sec:rotating_conversion}. The amplitude of the generated fast mode satisfies
\begin{equation}
    \delta E_F\propto \left(\frac{\omega_* r_m}{c}\right)\delta E_A.
\end{equation}
As a result, the conversion efficiency
\begin{equation}\label{eq:eff_omega}
    \frac{W_F}{W_0}\propto \left(\frac{\omega_* r_m}{c}\right)^2,
\end{equation}
independent of the \alfven wave amplitude. In Figure \ref{fig:eff_wrm} we plot the conversion efficiency at small wave amplitude as a function of $\omega_*r_m/c$. Indeed the trend roughly follows Equation (\ref{eq:eff_omega}). 
The three-wave interaction analysis in Appendix \ref{sec:rotating_conversion} is just for illustration; in reality the wavelength can be large and a full numerical treatment is needed.

The effect of rotation is important at small \alfven wave amplitude. At large amplitude, when the wave electric field becomes larger than the background rotation-induced electric field, rotation effect becomes subdominant, and we see the conversion efficiency falls back to the non-rotating trend, as shown in Figure \ref{fig:eff_dB_log}, especially the cyan and magenta trends.

Rotation induced linear coupling between \alfven mode and fast mode is most important near the separatrix, where the rotation time scale is comparable to the \alfven wave travel time scale. Figure \ref{fig:rotating-dw0.01r1_25-fields} shows an example where the \alfven wave is launched closer to the separatrix. We see that the initial \alfven wave generates a fast wave, which then produces new \alfven waves, extending the original \alfven wave all the way to the separatrix.

We also find that in the rotating magnetosphere, the conversion efficiency depends on the wavelength $\lambda_{\parallel}$ of the \alfven wave along the magnetic field line, as shown in Figure \ref{fig:eff_lambda}. This is likely because waves with different $\lambda_{\parallel}$ interact with different scales in the background variation. However, it does not depend on the total length of the wave train $cT$. This is different from the non-rotating case, confirming that the conversion mechanism is different. In this rotating case, We expect the conversion efficiency to drop at very small $\lambda_{\parallel}$ where WKB approximation is applicable; in the WKB regime the dispersion relations for the \alfven mode and fast mode do not intersect (Appendix \ref{sec:WKB}), so the conversion should be small. 
However, due to the very high resolution requirement, we are unable to reliably simulate cases with $\lambda_{\parallel}\ll r_*$. 

In summary, for \alfven waves with $\lambda_{\parallel}\approx5r_*$ and wave train length $cT\gtrsim0.2r_m$ propagating on field lines with a maximum radial extent $5r_*\lesssim r_{m}\lesssim 0.5r_{\rm LC}$, we find that the conversion efficiencies in a few asymptotic regimes are the following
\begin{align}\label{eq:scaling}
    \frac{W_F}{W_0}\approx
    \begin{cases}
    \displaystyle
    0.8\left(\frac{\omega_*r_m}{c}\right)^2, & 
    \displaystyle \left(\frac{\delta B}{B}\right)_{\rm eq}\ll\frac{\omega_*r_m}{c}\lesssim0.5,\\
    \displaystyle
    0.2\left(\frac{\delta B}{B}\right)_{\rm eq}^2, &
    \displaystyle
    \frac{\omega_*r_m}{c}\ll\left(\frac{\delta B}{B}\right)_{\rm eq}\lesssim 1.
    \end{cases}
\end{align}
The first branch is applicable to very small amplitude \alfven waves such that the rotation of the background magnetosphere is important; this is the relation from Figure \ref{fig:eff_wrm}. The second branch applies to relatively large amplitude \alfven waves; rotation effect becomes negligible and the scaling follows that in Figure \ref{fig:nonrotating-eff_dw_dB}. 

So far we have measured the conversion efficiency $W_F/W_0$ for the first passage of the \alfven wave through the equator. After the reflection from the stellar surface, \alfven wave can continue to convert to fast mode, but the efficiency becomes lower. This is because the \alfven wave gradually becomes dephased \citep{2020ApJ...897..173B}: the wave front is stretched and becomes increasingly oblique with respect to the background magnetic field, due to different lengths of neighboring field lines. Appendix \ref{sec:dephasing} shows some examples of the dependence of the conversion efficiency on the phase shift across the \alfven wave front. The conversion efficiency decreases as the phase shift increases, suggesting that conversion to fast mode requires the coherent $k_{\parallel}$ part of the \alfven wave.

\section{Implication for the Vela pulsar}\label{sec:Vela}
The Vela pulsar has a spin period of $P=2\pi/\omega_*=89$ ms, so the light cylinder is located at $R_{\rm LC}=c/\omega_*=4.2\times10^8$ cm. Taking the neutron star radius to be $r_*=10$ km, we have $R_{\rm LC}/r_*\approx4.2\times10^2$.
If the quake is triggered in the deep crust by a shear layer of thickness comparable to the local scale height, then the characteristic frequency of the waves is\footnote{The quake excites a broad spectrum of frequencies extending much higher than this characteristic frequency.} $\omega_A\sim10^4\,\rm{rad}\,\rm{s}^{-1}$ \citep{2020ApJ...897..173B}, and the corresponding wave length of the \alfven wave is $\lambda_A\sim 3\times10^6\,\rm{cm}$, a few times of $r_*$.
The wave train is likely long; for a quake duration of $T\sim100$ ms, the length of the launched wave train would be $3\times10^{9}$ cm.
The energy of the quake is not well constrained; for a rough estimation we take the characteristic energy flux of \alfven waves transmitted from the crust to the magnetosphere to be $F_*=10^{26}F_{*,26}\,\rm{erg}\,\rm{s}^{-1}\,\rm{cm}^{-2}$. The amplitude of the \alfven waves at the stellar surface is then $\delta B/B\approx10^{-4}F_{*,26}^{1/2}$.
If we simply follow the dipole scaling, the \alfven wave amplitude at a radius $r$ is $\delta B/B\approx 10^{-4}F_{*,26}^{1/2} (r/r_*)^{3/2}$. 
Thus, $\delta B/B\sim0.3$ at $r=0.5R_{\rm LC}$ and $\delta B/B\sim0.86$ at the light cylinder.
This can be marginally considered as a small amplitude \alfven wave, so the rotation effect is important in determining the efficiency of \alfven waves converting to fast modes. Applying the first branch in the scaling relation (\ref{eq:scaling}), we can see that if the \alfven wave is propagating on a flux tube that crosses the equator at half the light cylinder radius, the conversion efficiency after one pass is $\sim0.2$. 
If the \alfven wave is propagating closer to the separatrix, then the conversion efficiency can be higher, reaching $\sim0.3$. For waves with $(\delta B/B)_{\rm eq}\sim1$, using the second branch in the scaling relation (\ref{eq:scaling}), we get a conversion efficiency $\sim0.2$ as well. In these scenarios, the \alfven wave will 
lose a fraction $\gtrsim20\%$ of its initial energy during the first passage through the equator. Afterward the conversion efficiency decreases due to the dephasing of the wave, so the \alfven wave may keep bouncing in the magnetosphere for some time, until it loses most of its energy through this and other channels discussed in \S\ref{sec:intro}.

\section{Discussion and conclusion}\label{sec:conclusion}
In this paper we investigated the propagation of small amplitude \alfven waves in the closed zone of a dipolar pulsar magnetosphere. 
In the force-free regime \alfven waves can convert to fast magnetosonic waves as they propagate along curved field lines. We measured the conversion efficiency and obtained its scaling in different regimes (Equation \ref{eq:scaling}). The conversion efficiency is high for relatively large amplitude waves, and for waves propagating close to the separatrix/Y point, before the waves get significantly dephased. Typical \alfven waves launched by a quake in the Vela pulsar may convert to fast waves with an efficiency as high as 0.2 during the first passage, if the waves propagate to the outer region of the closed zone. However, the conversion efficiency decreases due to dephasing on subsequent passages. Therefore, during the $\sim 0.3$ seconds of quenched radio emission from Vela, the conversion to fast mode is not able to fully suppress the Alfven waves in the closed part of the magnetosphere. Thus we are currently unable to explain the short duration of the quenched radio emission during the glitch. This requires more detailed study of the quenching mechanism and other dissipation processes of the \alfven waves.

Similar processes could also happen in magnetar magnetospheres. Recently \citet{2020ApJ...900L..21Y} studied the fate of large amplitude \alfven waves launched by a magnetar quake. If the \alfven wave packet propagating on a flux tube with a radial extent $R$ has an energy larger than the magnetospheric energy $B^2R^3$ at $R$, the \alfven wave packet could break out from the magnetosphere and launch a relativistic ejecta. These may power X-ray bursts by particle acceleration in the current sheet behind the ejecta, and even produce fast radio bursts by masers at the shock \citep[e.g.,][]{1992ApJ...391...73G,2014MNRAS.442L...9L,2017ApJ...843L..26B,2020ApJ...896..142B,2019MNRAS.485.4091M,2019MNRAS.485.3816P,2020MNRAS.494.4627M,2020ApJ...899L..27M} or colliding plasmoids in the current sheet \citep{2019MNRAS.483.1731L,2019ApJ...876L...6P,2020ApJ...897....1L}. For smaller amplitude \alfven waves, the picture we studied in this paper applies. A moderate fraction of the \alfven wave energy could escape as it converts to fast waves; the rest of the wave energy may be dissipated in the magnetosphere through the channels discussed in \S\ref{sec:intro}.

In this paper we only studied axisymmetric modes. When the axisymmetry constraint is relaxed, more wave modes can participate in the interaction, which could change the conversion efficiency. Full 3D simulations are needed to quantify these effects.

Furthermore, in the force-free fluid framework, we are essentially considering the low frequency limit of the plasma modes. Kinetic effects may become important when the wavelength gets close to the plasma skin depth, or wave frequency becomes comparable to plasma frequency. In this regime, the \alfven waves may experience cutoff and resonance, and may undergo conversion to other plasma modes. This needs to be studied using a kinetic framework.

In our force-free simulations, we observe strong dephasing of \alfven waves, especially when the wave has passed through the equator and propagates back toward the star, consistent with \citet{2020ApJ...897..173B}. This leads to numerical dissipation. In reality, the strong shearing of the wave front could lead to a strong increase in the current density; this may trigger pair cascade or other types of plasma instability that dissipate away the \alfven wave energy. Kinetic simulations with physical dissipation mechanisms are required to study such processes and their influence on pulsar radio emission.

\acknowledgements
We thank Alex Chen, Xinyu Li and Anatoly Spitkovsky for helpful discussions. Y.Y. is supported by a Flatiron Research Fellowship at the Flatiron Institute, Simons Foundation. Y. L. and A. B. are supported by NSF grant 2009453 and by Simons Foundation grant 727992. A. P. is supported by NSF grant 1909458.

\software{{\it Coffee}, \url{https://github.com/fizban007/CoffeeGPU}, \citet{2020ApJ...893L..38C}}

\appendix
\twocolumngrid
\section{Calculation of the wave energy}\label{sec:wave_energy}
Suppose the background equilibrium has a magnetic field $\mathbf{B}_0$ and electric field $\mathbf{E}_0$. Writing the perturbation magnetic field as $\delta\mathbf{B}$ and perturbation electric field as $\delta\mathbf{E}$, we can calculate the wave energy from
\begin{align}\label{eq:wave-energy-full}
    W&=\int\frac{1}{2}[(\mathbf{B}_0+\delta \mathbf{B})^2+(\mathbf{E}_0+\delta \mathbf{E})^2]\, dV
    -\int\frac{1}{2}(\mathbf{B}_0^2+\mathbf{E}_0^2)\, dV\nonumber\\
    &=\int(\mathbf{B}_0\cdot\delta\mathbf{B}+\mathbf{E}_0\cdot\delta\mathbf{E})\, dV+\int\frac{1}{2}(\delta\mathbf{B}^2+\delta\mathbf{E}^2)\, dV
\end{align}
Making use of Maxwell equations, we find that the linear terms satisfy
\begin{align}\label{eq:rotating-energy-linear}
    &\frac{d}{dt}\int(\mathbf{B}_0\cdot\delta\mathbf{B}+\mathbf{E}_0\cdot\delta\mathbf{E})\, dV\nonumber\\
    &=\int[-\mathbf{B}_0\cdot(\nabla\times\delta\mathbf{E})+\mathbf{E}_0\cdot(\nabla\times\delta\mathbf{B}-\delta\mathbf{J})]\, dV\nonumber\\
    &=\oint (\mathbf{B}_0\times\delta\mathbf{E})\cdot d\mathbf{S}+\oint(\delta\mathbf{B}\times\mathbf{E}_0)\cdot d\mathbf{S}\nonumber\\
    &-\int(\mathbf{J}_0\cdot\delta\mathbf{E}+\mathbf{E}_0\cdot\delta\mathbf{J})\, dV\nonumber\\
    &=\oint(\delta\mathbf{B}\times\mathbf{E}_0)\cdot d\mathbf{S}-\int(\mathbf{J}_0\cdot\delta\mathbf{E}+\mathbf{E}_0\cdot\delta\mathbf{J})\, dV
\end{align}
where we have used $\nabla\times\mathbf{E}_0=0$ from the initial equilibrium condition, and $(\mathbf{B}_0\times\delta\mathbf{E})\cdot d\mathbf{S}=0$ from our stellar boundary condition  (\ref{eq:perturbation}-\ref{eq:perturbation_deltaE}). Similarly, the quadratic terms satisfy
\begin{align}\label{eq:rotating-energy-quadratic}
    &\frac{d}{dt}\int\frac{1}{2}(\delta\mathbf{B}^2+\delta\mathbf{E}^2)\, dV\nonumber\\
    &=\oint(\delta\mathbf{B}\times\delta\mathbf{E})\cdot d\mathbf{S}-\int\delta\mathbf{E}\cdot\delta\mathbf{J}\, dV.
\end{align}
The force-free constraint ensures $\mathbf{J}\cdot\mathbf{E}=0$, namely, $\mathbf{J}_0\cdot\delta\mathbf{E}+\mathbf{E}_0\cdot\delta\mathbf{J}+\delta\mathbf{E}\cdot\delta\mathbf{J}=0$, but keep in mind that this may not be enforced at each order. To sum up, we have
\begin{equation}\label{eq:dW_dt}
    \frac{dW}{dt}=\oint(\delta\mathbf{B}\times\mathbf{E}_0)\cdot d\mathbf{S}+\oint(\delta\mathbf{B}\times\delta\mathbf{E})\cdot d\mathbf{S}.
\end{equation}

If the background is a non-rotating dipole, $\mathbf{E}_0=0$ and $\mathbf{J}_0=0$, so equation (\ref{eq:rotating-energy-linear}) is identically zero. This means that the contribution to the wave energy comes purely from the quadratic terms, namely
\begin{equation}
    W=\int\frac{1}{2}(\delta\mathbf{B}^2+\delta\mathbf{E}^2)\, dV.
\end{equation}
Further more, after the perturbation, our line tying boundary condition ensures that $\delta\mathbf{E}$ is perpendicular to the stellar surface and $\delta\mathbf{B}$ is parallel to the stellar surface, so the right hand side of equation (\ref{eq:dW_dt}) is zero. This means that after the perturbation, the total energy of the system is conserved. 

When the background is a rotating, force-free dipole, the first term on the right hand side of equation (\ref{eq:rotating-energy-linear}) and (\ref{eq:dW_dt}) can be nonzero if $\delta\mathbf{B}_{\phi}\ne0$. For example, when an \alfven wave in the closed zone is launched or reflected at the stellar surface, $(\delta\mathbf{B}\times\mathbf{E}_0)\cdot d\mathbf{S}\ne0$, but after integrating over the full wave cycle, the change in $W$ contributed by this term can be zero. 
After the initial perturbation, the line tying boundary condition means that the second term in equation (\ref{eq:dW_dt}) is zero, so the total energy is conserved, in a time averaged sense.
The second term in equation (\ref{eq:rotating-energy-linear}) may not be zero, therefore the linear terms could have some contribution to the total wave energy. Nevertheless, for convenience we still use the quadratic terms $\int dV(\delta\mathbf{B}^2+\delta\mathbf{E}^2)/2$ as a measure of the wave energy. In Figure \ref{fig:rotating-dw0.01r1_10-energy}, the total quadratic energy slightly increases, which may be a consequence of the linear interaction between the \alfven wave and the background in the rotating case.


\section{Effect of numerical resolution}\label{sec:resolution}

\begin{figure}
    \centering
    \includegraphics[width=\columnwidth]{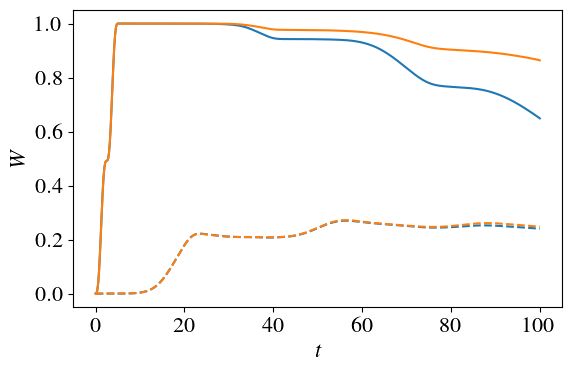}
    \caption{A comparison of energy history for the simulation shown in Figure \ref{fig:nonrotating-fields}, using 2 different resolutions. Solid lines are the total wave energies and dashed lines are fast wave energies. Orange lines correspond to a resolution of $6720\times4096$ in $(\log r, \theta)$ (the same as that shown in Figure \ref{fig:nonrotating-energy}), and blue lines correspond to a resolution of $3360\times2048$.}
    \label{fig:energy_res}
\end{figure}

In Figure \ref{fig:energy_res} we show a comparison of the energy history for different resolutions. Before the dephasing of the \alfven wave packet ($t\lesssim30r_*/c$), the total energy is well conserved in both resolutions. Around $t=(30-40)r_*/c$ and $t=(60-70)r_*/c$,
strong dephasing and spatial contraction of the \alfven wave near the stellar surface leads to numerical dissipation, and the dissipation is higher in the lower resolution run. On the other hand, the fast wave energy is almost identical in the two runs, indicating that the conversion to fast mode is physical and not dependent on numerical resolution.

\section{Effect of dephasing on the conversion efficiency} \label{sec:dephasing}
In order to investigate the effect of dephasing on the conversion efficiency in a controlled way, we carry out the following experiments. We introduce phase shift across $\theta$ at the launching of the wave, by modifying the perturbation (\ref{eq:perturbation}) into the following form
\begin{equation}\label{eq:perturbation_shear}
    \delta\omega=
    \begin{cases}
    \displaystyle
    \delta\omega_0e^{-\frac{1}{2}\left(\frac{\theta-\theta_m}{\sigma}\right)^2}\sin(2\pi n t_1/T), & 0\le t_1\le T,\\
    0, & t_1>T,
    \end{cases}
\end{equation}
where $t_1=t-s(\theta_2-\theta)/(\theta_2-\theta_1)$, and $s$ is introduced to be the initial phase shift between $\theta_1$ and $\theta_2$. Positive $s$ gives a phase shift that is in the same direction as would be generated by the propagation effect.

\begin{figure}
    \centering
    \includegraphics[width=\columnwidth]{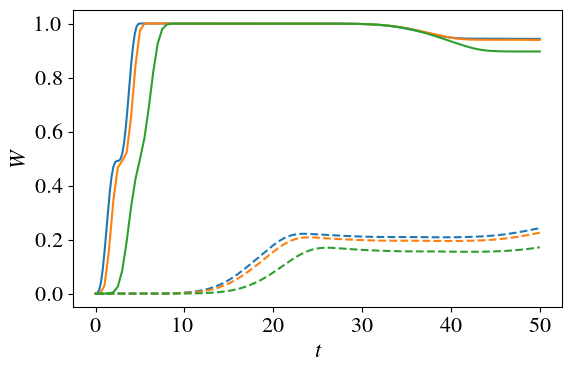}
    \caption{Energy history of runs in a non-rotating dipolar magnetosphere similar to Figure \ref{fig:nonrotating-energy}, but with different initial phase shift $s$ between the two boundaries of the wave packet at $\theta_1$ and $\theta_2$. Solid lines correspond to the total wave energy and dashed lines are the energy of the fast mode. Blue: $s=0$, orange: $s=1$, green: $s=5$.}
    \label{fig:energy_shear}
\end{figure}

Figure \ref{fig:energy_shear} shows one example of the wave energy evolution during the first passage of the \alfven wave through the equator in a non-rotating dipolar magnetosphere. We show the results of different phase shift $s$, with everything else fixed. Although the total energy of the injected wave packet is the same, we clearly see the drop of the conversion efficiency as $s$ increases. This in a way demonstrates that conversion to fast mode requires the coherent $k_{\parallel}$ part of the \alfven wave. When we introduce the phase shift, the wave packet becomes increasingly dominated by $k_{\perp}$ modes as $s$ increases, reducing the fraction that can convert to fast mode.

\begin{figure}
    \centering
    \includegraphics[width=\columnwidth]{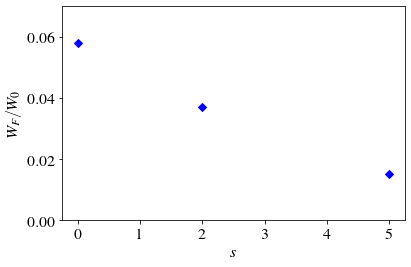}
    \caption{A comparison of conversion efficiency after the first pass, for \alfven waves launched with different initial phase shift $s$ in a rotating magnetosphere. The $s=0$ case is the same as that in Figures \ref{fig:rotating-dw0.01r1_10-fields} and \ref{fig:rotating-dw0.01r1_10-energy}.}
    \label{fig:eff_shear}
\end{figure}

We observe similar effect in a rotating magnetosphere. Figure \ref{fig:eff_shear} shows one example. This suggests that in a rotating magnetosphere, it is also the coherent $k_{\parallel}$ part of the \alfven wave that can convert to fast mode.

\section{WKB expansion of force-free normal modes} \label{sec:WKB}
Consider a steady state background with magnetic field $\mathbf{B}_0$ and electric field $\mathbf{E}_0$, satisfying $\nabla\times\mathbf{E}_0=0$, $\nabla\times\mathbf{B}_0=\mathbf{J}_0$. For wave modes whose wavelengths are much smaller compared to the length scale of background variation, we can make the WKB ansatz and write the wave electric field as $\delta\mathbf{E}\,e^{i\varphi}$. Plugging this into equations (\ref{eq:FF_dEdt}-\ref{eq:FF_J}) and linearize, we get the lowest order equation in the WKB expansion as
\begin{align}
    (\mathbf{k}\times(\mathbf{k}\times\delta\mathbf{E}))\cdot\left(\mathbf{I}-\frac{\mathbf{B}_0\mathbf{B}_0}{B_0^2}\right)+\omega^2\delta\mathbf{E}\nonumber\\
    -\frac{\omega}{B_0^2}\left[ (\mathbf{k}\cdot\delta \mathbf{E})\,\mathbf{E}_0\times\mathbf{B}_0-\mathbf{E}_0\cdot(\mathbf{k}\times\delta\mathbf{E})\mathbf{B}_0\right]=0,
\end{align}
where $\mathbf{k}=\nabla\varphi$ and $\omega=-\partial\varphi/\partial t$.
This is essentially the dispersion relation in a locally uniform $\mathbf{B}_0$ and $\mathbf{E}_0$ background. Suppose $\mathbf{B}_0$ is along $\hat{z}$ and $\mathbf{E}_0$ is along $\hat{x}$, we can obtain the following two normal modes: \\
(1) fast mode
\begin{align}
    \omega^2&=k^2,\\ 
    \delta\mathbf{E}&=C \left\{-\frac{E_0 \left(k_y^2+k_z^2\right)}{B_0k_xk_z}-\frac{\omega  k_y}{k_xk_z}, \frac{E_0 k_y}{B_0k_z}+\frac{\omega}{k_z} ,\frac{E_0 }{B_0}\right\}
\end{align}
(2) \alfven mode
\begin{align}
    \omega &=-\frac{E_0 k_y}{B_0}\pm \sqrt{1-\frac{E_0^2}{B_0^2}} k_z,\\
    \delta\mathbf{E}&=C \left\{k_x,\frac{E_0 \omega }{B_0}+k_y,\frac{E_0 k_z \left(B_0 k_y+E_0 \omega \right)}{B_0 \left(B_0 \omega +E_0 k_y\right)}\right\}.
\end{align}
The two branches of dispersion relations do not intersect with each other unless $\mathbf{k}$ is parallel to $\mathbf{B}_0$ in the comoving frame (the frame where $\mathbf{E}_0=0$).

In the closed zone of a rotating magnetosphere, both $\mathbf{B}_0$ and $\mathbf{E}_0$ are on the poloidal plane. Our 2D axisymmetry constraint means that $\mathbf{k}$ is also on the poloidal plane. Setting $k_y=0$ in the above expressions, we obtain\\
(1) fast mode
\begin{align}
    \omega^2&=k^2,\\ 
    \delta\mathbf{E}&=C \left\{-\frac{E_0 k_z}{B_0k_x}, \frac{\omega}{k_z} ,\frac{E_0 }{B_0}\right\}
\end{align}
(2) \alfven mode
\begin{align}
    \omega &=\pm \sqrt{1-\frac{E_0^2}{B_0^2}} k_z,\\
    \delta\mathbf{E}&=C \left\{k_x,\frac{E_0 \omega }{B_0},\frac{E_0^2 }{B_0^2}k_z\right\}.
\end{align}
With $\mathbf{E}_0\ne0$, the two branches of dispersion relations do not intersect in this case.
In order to investigate possible mode conversions, we need to go beyond the WKB approximation.

\section{Mode conversion in a rotating magnetosphere} \label{sec:rotating_conversion}
We illustrate the mode conversion in a varying background in the following simplified example. We assume that the background magnetic field is $\mathbf{B}=B_0\hat{z}+\mathbf{b}$, and the background electric field is $\mathbf{E}$. Both $\mathbf{b}$ and $\mathbf{E}$ are spatially varying. The background equilibrium has $\mathbf{J}=\nabla\times\mathbf{B}$, $\rho=\nabla\cdot\mathbf{E}$, and $\rho\mathbf{E}+\mathbf{J}\times\mathbf{B}=0$. We assume that $b$ and $E$ are small compared to $B_0$. The solutions to the force-free equations can then be obtained by asymptotic expansion in terms of $\mathbf{b}$. Our derivation is in a way similar to the three-wave interaction process discussed by \citet{2019MNRAS.483.1731L}.

In zeroth order of $b$, we obtain the usual normal modes in a uniform magnetic field. Using $\delta\mathbf{E}_k^{(0)}$ to denote the amplitude of the wave electric field for wave vector $k$, we can write the modes as\\
(1) fast mode
\begin{align}
    \delta\mathbf{E}_k^{(0)}&=\frac{ \hat{z}\times \mathbf{k}}{ k \sin \theta}\delta E_k^{(0)},\\
    \delta\mathbf{B}_k^{(0)}&=\frac{ \left(k \hat{z}- \mathbf{k} \cos\theta  \right)}{ k \sin\theta}\delta E_k^{(0)},\\
    \delta\rho_k^{(0)}&=0,\quad \delta\mathbf{J}_k^{(0)}=0.
\end{align}
(2) \alfven mode
\begin{align}
    \delta\mathbf{E}_k^{(0)}&=\frac{\mathbf{k}- \hat{z}k\cos\theta}{ k \sin \theta}\delta E_k^{(0)},\\
    \delta\mathbf{B}_k^{(0)}&=-\frac{ \mathbf{k}\times\hat{z}\cos\theta}{ \omega \sin\theta}\delta E_k^{(0)},\\
    \delta\rho_k^{(0)}&=i k\sin\theta \delta E_k^{(0)},\\
    \delta\mathbf{J}_k^{(0)}&={\rm sgn}(\cos\theta)\delta \rho_k^{(0)}\hat{z}.
\end{align}
To first order of $b$, we assume that the wave amplitude of the zeroth order solution slowly varies with time. So the Maxwell equations (\ref{eq:FF_dEdt}-\ref{eq:FF_dBdt}) become
\begin{align}
    i k\times \delta \mathbf{E}_k^{(1)} &= i \omega  \delta \mathbf{B}_k^{(1)}-\frac{\partial \delta \mathbf{B}_k^{(0)}}{\partial t},\label{eq:pert_dBdt}\\
    i k\times \delta \mathbf{B}_k^{(1)}&=\delta \mathbf{J}_k^{(1)}-i \omega  \delta \mathbf{E}_k^{(1)}+\frac{\partial }{\partial t}\delta \mathbf{E}_k^{(0)}.\label{eq:pert_dEdt}
\end{align}
The force-free constraint (\ref{eq:FF_constraint}) at this order becomes
\begin{align}
    \sum_{k'}\left(\rho_{k'}\delta\mathbf{E}_{k-k'}^{(0)}+\mathbf{J}_{k'}\times\delta\mathbf{B}_{k-k'}^{(0)}\right)
    +\delta\mathbf{J}_k^{(1)}\times\mathbf{B}_0=0,
\end{align}
where $\rho_{k'}$ and $\mathbf{J}_{k'}$ are the Fourier components of the background charge and current density. This is similar to a three-wave interaction process: a $\mathbf{k}_1$ wave could interact with the background $\mathbf{k}_2$ component (purely spatial) and generate a $\mathbf{k}$ wave if the resonant condition is satisfied: $\mathbf{k}_1+\mathbf{k}_2=\mathbf{k}$, $\omega_1=\omega$. Consider a single $\mathbf{k}_1$ mode to begin with, we have
\begin{align}\label{eq:pert_constraint}
    \rho_{k_2}\delta\mathbf{E}_{k_1}^{(0)}+\mathbf{J}_{k_2}\times\delta\mathbf{B}_{k_1}^{(0)}+\delta\mathbf{J}_k^{(1)}\times\mathbf{B}_0=0.
\end{align}
From Equations (\ref{eq:pert_dBdt}) and (\ref{eq:pert_dEdt}) we can obtain
\begin{align}
    \delta\mathbf{J}_k^{(1)}&=\frac{i}{\omega } \left(\left(\mathbf{k}\cdot \delta \mathbf{E}_k^{(1)}\right)\mathbf{k}+\left(\omega ^2-k^2\right)\delta \mathbf{E}_k^{(1)}\right)\nonumber\\
    &+\frac{1}{\omega ^2}\left(\mathbf{k}\cdot \frac{\partial \delta \mathbf{E}_k^{(0)}}{\partial t}\mathbf{k}-(k^2+\omega ^2)\frac{\partial \delta \mathbf{E}_k^{(0)}}{\partial t}\right)
\end{align}
This can then be plugged into (\ref{eq:pert_constraint}) to eliminate $\delta\mathbf{J}_k^{(1)}$. Now suppose $\mathbf{k}_1$ corresponds to an Alfven wave, and $\mathbf{k}$ corresponds to a fast wave. Using the zeroth order wave solutions, we obtain
\begin{align}
    \delta E_{k_1}^{(0)}\left(\rho_{k_2}\frac{\mathbf{k}_1-\hat{z}k_1\cos\theta_1}{k_1\sin\theta_1}-\mathbf{J}_{k_2}\times\frac{\mathbf{k}_1\times\hat{z}\cos\theta_1}{\omega_1\sin\theta_1}\right)\nonumber\\
    +\left(\frac{i}{\omega}\left(\mathbf{k}\cdot\delta \mathbf{E}_{k}^{(1)}\right)\mathbf{k}-2\frac{\hat{z}\times\mathbf{k}}{k\sin\theta}\frac{\partial \delta E_{k}^{(0)}}{\partial t}\right)\times\mathbf{B}_0=0.
\end{align}
Taking a dot product with $\mathbf{k}$, we get
\begin{align}
    \frac{\partial \delta E_{k}^{(0)}}{\partial t}&=\frac{\delta E_{k_1}^{(0)}}{2B_0k\sin\theta}\left(\rho _{k_2}\frac{ \mathbf{k}_1\cdot \mathbf{k}-\hat{z}\cdot \mathbf{k} \, k_1 \cos\theta_1  }{ k_1 \sin\theta_1 }\right.\nonumber\\
    &\left.-\mathbf{J}_{k_2}\cdot\frac{(\mathbf{k}_1\cdot\mathbf{k}\,\hat{z}-\mathbf{k}\cdot\hat{z}\,\mathbf{k}_1){\rm sgn}(\cos\theta_1)}{k_1\sin\theta_1}\right)
\end{align}
In analogy with a local region in the closed zone of an aligned rotator, we have $\mathbf{b}$ and $\mathbf{E}$ both on the poloidal plane. Suppose this is the $x-z$ plane. In our axisymmetric simulations both $\mathbf{k}_1$ and $\mathbf{k}$ are in $x-z$ plane as well. Background $\rho\ne0$ and $\mathbf{J}$ is along toroidal direction ($y$ direction). So the second term in the parentheses of the above equation is zero, and we have
\begin{align}
    \frac{\partial \delta E_{k}^{(0)}}{\partial t}=\frac{\delta E_{k_1}^{(0)}}{2B_0k\sin\theta}\rho _{k_2}\frac{ \mathbf{k}_1\cdot \mathbf{k}-\hat{z}\cdot \mathbf{k} \, k_1 \cos\theta_1  }{ k_1 \sin\theta_1 }.
\end{align}
So we can see that the growth of the outgoing fast mode amplitude is proportional to the amplitude of the incident \alfven mode, and the background charge density (on the appropriate scale).
Since close to the star, the background charge density is $\rho\approx \pmb{\omega}_*\cdot\mathbf{B}/(2\pi c)$, we see that $\partial\delta E_k^{(0)}/\partial t\propto \delta E_{k_1}^{(0)}\omega_*$. Waves propagating on a flux tube with a maximum radial extent $r_m$ have a typical travel time $t\sim r_m/c$, so the final outgoing fast mode amplitude satisfies $\delta E_k^{(0)}\propto (\omega_*r_m/c)\delta E_{k_1}^{(0)}$.


\bibliography{ref}
\bibliographystyle{aasjournal}


\end{document}